\documentclass[letter,reqno]{amsart}%

\usepackage{amsmath}
\usepackage{amssymb}
\usepackage{pb-diagram}
\usepackage{amsfonts}

\newtheorem{theorem}{Theorem}
\theoremstyle{plain}
\newtheorem{corollary}{Corollary}
\newtheorem{definition}{Definition}
\newtheorem{example}{Example}

\newtheorem{proposition}{Proposition}
\newtheorem{remark}{Remark}
\numberwithin{equation}{section}

\begin{document}

\title{Generating operators of the Krasil'shchik-Schouten bracket}
\author{Jos\'{e} A. Vallejo}
\address{Dept. Geometria i Topologia\\
Universitat de Val\`{e}ncia (Spain)\\
Postal Address: Av. V. A. Estell\'{e}s 1, 46100 Burjassot.\\
Facultad de Ciencias\\
Universidad Aut\'{o}noma de San Luis Potos\'{\i} (M\'{e}xico)\\
Postal Address: Lat. Av. Salvador Nava s/n, 78290 SLP.}
\email{jose.a.vallejo@fc.uaslp.mx}
\thanks{Acknowledgements: The author wants to express his gratitude to Mme. Yvette
Kosmann-Schwarzbach, for encouragement, careful reading, useful corrections
and comments. Also thanks are due to Josef Krasil'shchik for a thorough
revision and comments, and for pointing out the existence of the paper by
Grabowski and Marmo cited in the references.}
\date{December 18 2003}

\begin{abstract}
It is proved that given a divergence operator on the structural sheaf of
graded commutative algebras of a supermanifold, it is possible to construct a
generating operator for the Krashil'shchik-Schouten bracket. This is a
particular case of the construction of generating operators for a special
class of bigraded Gerstenhaber algebras.

Also, some comments on the generalization of these results to the context of
$n-$graded Jacobi algebras are included.

\end{abstract}


\maketitle

\section{Introduction}

These notes intend to describe, in a relatively self-contained manner, the
construction of a generating operator for the Krasil'shchik-Schouten bracket
on the multiderivations of a supermanifold $(M,\mathcal{A})$, $\mathcal{D}%
(\mathcal{A})$, once a divergence operator on the structural sheaf
$\mathcal{A}$ is given. For simplicity, we do not study the more general case,
in which $\mathcal{A}$ is a sheaf of $n-$graded commutative algebras, but we
restrict ourselves to the common situation in supermanifold theory of
$\mathbb{Z}-$graded commutative algebras (that is, $n=1$). Nevertheless, the
changes needed to deal with the general case are mainly of notational character.

Thus, we answer affirmatively a conjecture raised by Kosmann-Schwarzbach and
Monterde in \cite{Kos-Mon 02}. To put it in context, let us recall that the
notion of generating operator (and related Gerstenhaber algebras) has its
origins in the quantization of non abelian gauge theories through the
Batalin-Vilkovisky (BV) formalism (see \cite{Bat-Vil 81}, \cite{Kos 95b},
\cite{Wit 90}). Concretely, Witten showed in \cite{Wit 90} (see also \cite{Sch
93}) that the master equation and the condition of nilpotency of the BV
operator have an algebraic background: they make sense each time one has a
Gersternhaber algebra with generating operator of square zero. This is the
case, for example, of an odd Poisson structure and the associated Hamiltonian
divergence; a detailed study of these structures is presented in \cite{Kos-Mon
02}, where the analogous results in the graded case (the
Krasil'shchik-Schouten bracket) are left as a conjecture.

The first part of the paper contains a brief review of the basic notions of
supermanifold theory needed to understand the construction of the
Krasil'shchik-Schouten bracket on multiderivations, which is an abstract
generalization of the graded Schouten bracket on graded multivectors. After
presenting Krasil'shchik work, we introduce the notion of divergence operator,
a graded reformulation of the usual divergence in manifold theory. It is shown
how to extend such a divergence to a differential operator on the algebra of
multiderivations $\mathcal{D}(\mathcal{A})$, through a recursive algorithm,
and we prove that the result is a generating operator of the
Krashil'shchick-Schouten bracket on $\mathcal{D}(\mathcal{A})$. All the
construction is based on the fact that the algebra of multiderivations endowed
with the Krashil'shchick-Schouten bracket, $(\mathcal{D}(\mathcal{A}%
),[\![\_,\_]\!])$, is a bigraded Gerstenhaber algebra. The procedure to define
such generating operators is well known in the $\mathbb{Z}-$graded case, and
goes back to the work of Koszul \cite{Koz 85} (see also \cite{Kos 95} for a
study in the context of Lie algebroids); it turns out that the same ideas can
be applied in the bigraded context.

Finally, we make some digressions about the generalization of these results to
the case in which one deals with an $n-$graded Jacobi algebra, where Leibniz's
rule does not hold.

All the gradings, unless otherwise specified, will be understood as
$\mathbb{Z}-$gradings.

\section{Some results from Supermanifold Theory}

\subsection{Basic definitions}

For references, see \cite{Kos 77} or \cite{Lei 80}. Let $K$ be the commutative
field $\mathbb{R}$ or $\mathbb{C}$, and $M$ a differential manifold. The set
of differentiable functions on $M$ gives us an example of sheaf $\mathcal{F}$;
if $U$ is open in $M$, we have%
\[
\mathcal{F}(U)=\{f:U\rightarrow\mathbb{R}:f\in C^{\infty}(U)\}\equiv
C^{\infty}(U)
\]
and, for $V\subset U$ open,%
\begin{gather*}
\rho_{V}^{U}:C^{\infty}(U)\rightarrow C^{\infty}(V)\\
f\mapsto f|_{V}.
\end{gather*}

Indeed, this is a sheaf of graded commutative algebras, although in this case
the grading is trivial. The basic idea underlying the definition of a graded
manifold, is the substitution of the commutative sheaf of differentiable
functions by another one in which we can accommodate objects with a
$\mathbb{Z}_{2}-$grading.

\begin{definition}
A graded manifold (or supermanifold) of dimension $(m|n)$ and basis
$(M,C^{\infty}(M))$ is given by a usual differential manifold $M$, with
dimension $m$, and a sheaf $\mathcal{A}$ of graded $K-$commutative algebras
(the structural sheaf) such that

\begin{enumerate}
\item We have an exact sequence of sheaves%
\begin{equation}
0\rightarrow\mathcal{N}\rightarrow\mathcal{A}\overset{\sim}{\rightarrow
}C^{\infty}(M)\rightarrow0, \label{eq1}%
\end{equation}
where $\mathcal{N}$ is the sheaf of nilpotents of $\mathcal{A}$ and $\sim$ is
a surjective morphism of graded $K-$commutative algebras.

\item $\mathcal{N}/\mathcal{N}^{2}$ is a locally free module with rank $n$
over $C^{\infty}(M)=\mathcal{A}/\mathcal{N}$, and $\mathcal{A}$ is locally
isomorphic, as a sheaf of graded $K-$commutative algebras, to the exterior
bundle $\Lambda_{C^{\infty}(M)}(\mathcal{N}/\mathcal{N}^{2})$.
\end{enumerate}
\end{definition}

\begin{remark}
In the following, we will consider only real graded manifolds, that is, those
for which $K=\mathbb{R}$. Thus, we will omit the corresponding prefix.
\end{remark}

From the exact sequence (\ref{eq1}) we obtain, for any open $U\subset X,$ the
exact sequence of graded algebras%
\[
0\rightarrow\mathcal{N}(U)\rightarrow\mathcal{A}(U)\overset{\sim}{\rightarrow
}C^{\infty}(U)\rightarrow0.
\]
A section $f$ of $\mathcal{A}$ will be called a graded function (or
superfunction). The image of such a graded function $f\in\mathcal{A}(U)$ by
the structural morphism $\sim$ is denoted $\tilde{f}$.

Next, we introduce morphisms between graded manifolds.

\begin{definition}
A morphism of graded manifolds is a pair of mappings $(f,\phi):(M,\mathcal{A}%
)\rightarrow(N,\mathcal{B})$ where $f:M\rightarrow N$ is a differential
mapping between usual manifolds and for each open $U\subset N$, $\phi
:\mathcal{B}(U)\rightarrow\mathcal{A}(f^{-1}(U))$ is an even morphism of
graded algebras compatible with the restrictions, and all such that the
diagram%
\[
\begin{diagram} \node{\mathcal{B}(U)} \arrow{e,t}{\phi} \arrow{s} \node{\mathcal{A}(f^{-1}(U))} \arrow{s} \\ \node{C^{\infty}(U)} \arrow{e,b}{f_{\ast}} \node{f_{\ast}C^{\infty}(f^{-1}(U))} \end{diagram}
\]
commutes.
\end{definition}

Isomorphisms, as a particular case, are defined in an obvious way.

According to the definition, the structural sheaf of a supermanifold
$(M,\mathcal{A})$ (with dimension $(m|n)$) is locally isomorphic to the
locally free $C^{\infty}(M)-$module $\mathcal{N}/\mathcal{N}^{2}$, which has
rank $n $, so by the Serre duality theorem (see \cite{Ser 55}) there exists a
vector bundle of rank $n$ over $M$, $E$, such that any point of $M$ has an
open neighborhood $U$ for which $\mathcal{A}(U)\simeq\Gamma_{U}(\Lambda E)$ as
graded commutative $\mathbb{R}-$algebras. The splitting neighborhoods of a
supermanifold are those for which these two representations of the structural
sheaf are identified.

\begin{definition}
A splitting neighborhood of a supermanifold $(M,\mathcal{A})$ is an open
$U\subset M$ such that $E|_{U}$ is a trivial bundle and $\mathcal{A}%
|_{U}\simeq\Lambda_{C^{\infty}(U)}(\mathcal{N}/\mathcal{N}^{2})$.
\end{definition}

If $U$ is a splitting neighborhood for $(M,\mathcal{A})$ there exists a basis,
which we will denote $\{x^{-1},...,x^{-n}\}$, of sections of $E|_{U}$, along
with an isomorphism%
\[
\mathcal{A}(U)\simeq C^{\infty}(U)\otimes_{\mathbb{R}}\Lambda E_{n}%
\]
where $E_{n}=\left\langle \{x^{-1},...,x^{-n}\}\right\rangle $ is the vector
$\mathbb{R}-$space generated by $\{x^{-1},...,x^{-n}\}$. Also, there exists a
section $\sigma:C^{\infty}(U)\hookrightarrow\mathcal{A}(U)$ for the structural
morphism $\sim$.

\begin{definition}
If $U$ is a splitting neighborhood, a family $\{x^{i},x^{-j}\}_{1\leq i\leq
m}^{1\leq j\leq n}$ of superfunctions $(\left\vert x^{i}\right\vert
=0,\left\vert x^{-j}\right\vert =1)$ is called a graded coordinate system (or
supercoordinate system) if

\begin{enumerate}
\item $x^{i}=\sigma(\tilde{x}^{i})$ $(1\leq i\leq m),$ where $\{\tilde{x}%
^{1},...,\tilde{x}^{m}\}$ is an ordinary coordinate system on $U$.

\item $\{x^{-1},...,x^{-n}\}$ is a basis of sections of $E|_{U}$, that is,
$x^{-1},...,x^{-n}\in\Lambda E_{n}$ and $\prod\limits_{j=1}^{n}x^{-j}\neq0$.
\end{enumerate}
\end{definition}

\begin{remark}
As we have already mentioned, the structural sheaf of a supermanifold
$(M,\mathcal{A})$ is locally isomorphic to $\Lambda_{C^{\infty}(M)}%
(\mathcal{N}/\mathcal{N}^{2})$. An important Theorem (known as the Batchelor
theorem \cite{Bat 79}, but see also \cite{Gaw 77}), guarantees that this is
not true just locally, but also globally. Thus, for any supermanifold
$(M,\mathcal{A}) $ there exists a vector bundle over $M\,$, $E$, such that the
sheaf $\mathcal{A}$ can be identified with the sheaf of sections of the
exterior bundle associated to $E$. However, this isomorphism is not canonical.
\end{remark}

\subsection{Graded vector fields and differential forms}

\begin{definition}
Let $\mathcal{F},\mathcal{G}$ be sheaves on a topological space $X$. For any
open $U\subset M$, let $\mathrm{Hom}(\mathcal{F}|_{U},\mathcal{G}|_{U})$
denote the space of morphisms between the restricted sheaves $\mathcal{F}%
|_{U}$ and $\mathcal{G}|_{U}$. The sheaf of homomorphisms is the sheaf
$\mathrm{Hom}(\mathcal{F},\mathcal{G})$ given by $\mathrm{Hom}(\mathcal{F}%
,\mathcal{G})(U)=\mathrm{Hom}(\mathcal{F}|_{U},\mathcal{G}|_{U})$ with the
natural restriction morphisms.
\end{definition}

Let us consider now a graded manifold $(M,\mathcal{A})$, where $\mathcal{A}$
is a sheaf of graded commutative $\mathbb{R}-$algebras. Write End$_{\mathbb{R}%
}(\mathcal{A})=\mathrm{Hom}_{\mathbb{R}}(\mathcal{A},\mathcal{A})$ for the
sheaf of $\mathbb{R-}$linear endomorphisms.

\begin{definition}
The sheaf of derivations of $(M,\mathcal{A})$ is the subsheaf of
\textrm{End}$_{\mathbb{R}}(\mathcal{A})$ whose sections on an open subset
$U\subset M$ are $\mathbb{R}-$linear graded derivations $D:\mathcal{A}%
|_{U}\rightarrow\mathcal{A}|_{U}$. This sheaf is denoted \textrm{Der}%
$_{\mathbb{R}}(\mathcal{A})$ or simply \textrm{Der}$(\mathcal{A})$, and its
elements are called the graded vector fields (or supervector fields) of the
graded manifold $(M,\mathcal{A})$. Sometimes, we will write \textrm{Der}%
$(\mathcal{A})=\mathcal{X}_{G}(M)$.
\end{definition}

Let $U$ be a coordinate neighborhood for a graded manifold $(M,\mathcal{A})$
with graded coordinates $\{x^{i},x^{-j}\}_{1\leq i\leq m}^{1\leq j\leq n}$.
There exist even derivations $\frac{\partial}{\partial x^{1}},...,\frac
{\partial}{\partial x^{m}}$ and odd derivations $\frac{\partial}{\partial
x^{-1}},...,\frac{\partial}{\partial x^{-m}}$ of $\mathcal{A}(U)$ uniquely
characterized by the conditions%
\[%
\begin{array}
[c]{cccc}%
\frac{\partial x^{j}}{\partial x^{i}}=\delta_{i}^{j}; & \frac{\partial x^{-j}%
}{\partial x^{i}}=0; & \frac{\partial x^{j}}{\partial x^{-i}}=0; &
\frac{\partial x^{-j}}{\partial x^{-i}}=\delta_{i}^{j}%
\end{array}
\]
(negative indices run from $-n$ to $-1$, positive ones from $1$ to $m$) and
such that every derivation $X\in$\textrm{Der}$(\mathcal{A}(U))$ can be written
as%
\[
X=\sum\limits_{i=1}^{m}X(x^{i})\frac{\partial}{\partial x^{i}}+\sum
\limits_{j=1}^{m}X(x^{-j})\frac{\partial}{\partial x^{-j}},
\]
where $X(f)$ denotes the acion of $X$ on a superfunction. In particular,
\textrm{Der}$(\mathcal{A}(U))$ is a free right $\mathcal{A}(U)-$module with
basis $\frac{\partial}{\partial x^{1}},...,\frac{\partial}{\partial x^{m}}; $
$\frac{\partial}{\partial x^{-1}},...,\frac{\partial}{\partial x^{-m}}$.

If $U\subset M$ is an open set, the algebraic dual of the graded
$\mathcal{A}(U)-$module \textrm{Der}$(\mathcal{A}(U))$ is \textrm{Der}$^{\ast
}(\mathcal{A}(U))=\mathrm{Hom}_{\mathcal{A}}($\textrm{Der}$(\mathcal{A}%
(U)),\mathcal{A}(U))$, which has itself a natural structure of graded
$\mathcal{A}(U)-$module and we get then a sheaf $U\rightarrow$\textrm{Der}%
$^{\ast}(\mathcal{A}(U))$. On \textrm{Der}$^{\ast}(\mathcal{A})$ we can
introduce an exterior product in the usual way, obtaining the following.

The sheaves of graded homogeneous differential forms on $(M,\mathcal{A})$ are
the sheaves%
\[
\Omega_{G}^{p}(\mathcal{A})=\Lambda^{p}\mathrm{Der}^{\ast}(\mathcal{A}).
\]
We will put $\Omega_{G}(\mathcal{A})=\sum\limits_{p\in\mathbb{N}}\Omega
_{G}^{p}(\mathcal{A})$ (also, we understand $\Omega_{G}^{0}(\mathcal{A}%
)=\mathcal{A}$).

The graded differential forms on $(M,\mathcal{A})$ will be sometimes called
graded forms.

Being a graded homomorphism of graded modules, a graded differential form has
a degree. Thus, we can define a $\mathbb{Z}\times\mathbb{Z}-$bigrading on the
module of graded differential forms and we will say that a graded homogeneous
differential form $\lambda$ has bidegree $(p,k)\in\mathbb{Z}\times\mathbb{Z}$
(denoted $\lambda\in\Omega_{G}^{(p,k)}(\mathcal{A})$) if
\[
\lambda:\mathrm{Der}(\mathcal{A})\times.\overset{p)}{.}.\times\mathrm{Der}%
(\mathcal{A})\longrightarrow\mathcal{A}%
\]
and if, for all $X_{1},...,X_{p}\in$ \textrm{Der}$(\mathcal{A})$,
\[
\left\vert \left\langle X_{1},...,X_{p};\lambda\right\rangle \right\vert
=\sum\limits_{i=1}^{p}\left\vert X_{i}\right\vert +k\text{.}%
\]

Thus, we have $\Omega_{G}^{p}(\mathcal{A})=\sum\limits_{k\in\mathbb{Z}}%
\Omega_{G}^{(p,k)}(\mathcal{A})$.

\begin{remark}
In order to avoid confusions when speaking about bigraded objects, we adopt
the convention of naming \textquotedblleft cohomological
degree\textquotedblright\ the degree which indicates the number of arguments
that the object admits ($p$ above), and \textquotedblleft ghost
degree\textquotedblright\ the other one (this terminology comes from physics,
more precisely from the quantization of non abelian gauge theories).
\end{remark}

Using this bigrading, any arbitrary (of non homogeneous ghost degree) graded
$p-$differential form $\lambda$ can be decomposed as a sum $\lambda
=\lambda_{(0)}+...+\lambda_{(n)}$, where $\lambda_{(i)}$ is an homogeneous
graded form of bidegree $(p,i)$.

The insertion operator is defined as usual. If $\lambda$ is a graded form of
bidegree $(p,k)$ and $X$ is a derivation of degree $\left\vert X\right\vert $,
then $\iota_{X}\lambda$ is the graded form of bidegree $(p-1,k+\left\vert
X\right\vert )$ defined by
\[
\langle X_{1},\dots,X_{p-1};\iota_{D}\lambda\rangle:=\langle X_{1}%
,\dots,X_{p-1},X;\lambda\rangle.
\]

This implies that the bidegree of the operator $\iota(X)=\iota_{X}$ is
$(-1,|X|)$.

We shall denote by $\operatorname{d}^{G}$ the graded exterior differential.
(See \cite{Kos 77} for details.) In particular, for a graded $0$-form
$\alpha\in\mathcal{A}$, $\langle X;\operatorname{d}^{G}\alpha\rangle
=X(\alpha)$, and for a graded $1$-form, $\lambda$, on $(M,\mathcal{A})$ we
have
\[
\langle X_{1},X_{2};(\operatorname{d}^{G}\lambda)\rangle=X_{1}(\langle
X_{2};\lambda\rangle)-(-1)^{\left\vert X_{1}\right\vert \left\vert
X_{2}\right\vert }X_{2}(\langle X_{1};\lambda\rangle)-\langle\lbrack
X_{1},X_{2}];\lambda\rangle.
\]
The graded exterior differential is an operator of bidegree $(1,0)$.

Other familiar operators on ordinary manifolds also have counterparts on
graded manifolds. If $X\in\mathrm{Der}(\mathcal{A})$, then the Lie operator
$\mathcal{L}_{X}^{G}$ is defined by
\begin{equation}
\mathcal{L}_{X}^{G}=\iota_{X}\circ\operatorname{d}^{G}+\operatorname{d}%
^{G}\circ\iota_{X}. \label{eq1b}%
\end{equation}

Note that $\mathcal{L}_{X}^{G}$ is a derivation of bidegree $(0,|X|)$.

\subsection{Graded multivector fields}

We have introduced in the preceding Section the concept of graded vector
field; now, we study graded multivector fields.

Let $\Omega_{G}^{p}(\mathcal{A})=\sum\limits_{k\in\mathbb{Z}}\Omega
_{G}^{(p,k)}(\mathcal{A})$. A graded $q-$multivector field $A$ is a
$C^{\infty}(M)-$ multilinear alternating morphism of sheaves
\[
A:\Omega_{G}^{1}(\mathcal{A})\times.\overset{q)}{.}.\times\Omega_{G}%
^{1}(\mathcal{A})\longrightarrow\mathcal{A},
\]
and, just as in the case of graded differential forms, the graded multivector
fields are actually bigraded objects. We will say that a graded multivector
field $A$ has bidegree $(A_{1},A_{2})\in\mathbb{Z}\times\mathbb{Z}$ (and will
denote it $A\in\Gamma(\Lambda_{G}^{(A_{1},A_{2})}(\mathcal{A}))$ or $A\in
A_{G}^{(A_{1},A_{2})}(\mathcal{A})$) if
\[
\left\vert A(\lambda_{1},...,\lambda_{A_{1}})\right\vert =\sum\limits_{i=1}%
^{q}\left\vert \lambda_{i}\right\vert +A_{2};
\]
also we denote $\Gamma(\Lambda_{G}^{q}(\mathcal{A}))=\sum\limits_{k\in
\mathbb{Z}}\Gamma(\Lambda_{G}^{(q,k)}(\mathcal{A}))$.

Graded multivector fields are generated by decomposable elements of the form
\[
A=X_{1}\wedge...\wedge X_{q}\in\Gamma(\Lambda_{G}^{q}(\mathcal{A})),
\]
where $X_{i}\in\mathrm{Der}(\mathcal{A})$, $1\leq i\leq q$.

Associated to each $A\in\Gamma(\Lambda_{G}^{q}(\mathcal{A}))$, we have the
insertion operator $\iota_{A}$, which is defined in analogy with the
corresponding operator for graded vector fields; if $A$ is a decomposable
element as above, we have
\begin{align*}
\iota_{A}  &  :\Omega_{G}^{p}(\mathcal{A})\rightarrow\Omega_{G}^{p-q}%
(\mathcal{A})\\
\varsigma &  \mapsto\iota_{X_{q}}\circ...\circ\iota_{X_{1}}(\varsigma)\text{.}%
\end{align*}

Note that $\iota_{A}$ is an operator on $\Omega_{G}(\mathcal{A})=\sum
\limits_{p\in\mathbb{Z}}\Omega_{G}^{p}(\mathcal{A})$ with bidegree
$(-q,\sum\limits_{i=1}^{q}X_{i})$ (for ease in writing, we will omit the bars
when denoting degrees, if there is no risk of confusion). In the same way, the
operator $\mathcal{L}_{A}^{G}=\iota_{A}\circ\operatorname{d}^{G}%
+\operatorname{d}^{G}\circ\iota_{A}$ has bidegree $(-q+1,\sum\limits_{i=1}%
^{q}X_{i})$. It generalizes the graded Lie derivative (\ref{eq1b}).

\subsection{The graded Schouten-Nijenhuis bracket\label{GSN}}

We have already found several examples of differential operators. Let us
consider them in more detail.

\begin{remark}
In the following, we will particularize to differential operators on
$\Omega_{G}(\mathcal{A})$, because this is the main example we will use. But
all the definitions and properties remain unchanged, with the obvious
modifications in notation, if we consider a general bigraded (or $n-$graded)
algebra $\mathcal{B}$ instead of $\Omega_{G}(\mathcal{A})$.
\end{remark}

Let $\mathrm{End}_{\mathbb{R}}(\Omega_{G}(\mathcal{A}))$ be the space of
$\mathbb{R}-$endomorphisms of the sheaf of graded algebras $\Omega
_{G}(\mathcal{A})$. We say that $\Delta\in\mathrm{End}_{\mathbb{R}}(\Omega
_{G}(\mathcal{A}))$ has bidegree $(\Delta_{1},\Delta_{2})$ if $\Delta$ maps
$\Omega_{G}^{(p,q)}(\mathcal{A})$ to $\Omega_{G}^{(p+\Delta_{1},q+\Delta_{2}%
)}$, and if this is the case we write $\Delta\in\mathrm{End}_{\mathbb{R}%
}^{(\Delta_{1},\Delta_{2})}(\Omega_{G}(\mathcal{A}))$.

On the space $\mathrm{End}_{\mathbb{R}}(\Omega_{G}(\mathcal{A}))$ we introduce
a bracket $[.,.]$ (called \ graded commutator) by means of
\[
\lbrack\Delta,\Gamma]=\Delta\circ\Gamma-(-1)^{\left\langle \Delta
,\Gamma\right\rangle }\Gamma\circ\Delta\in\mathrm{End}_{\mathbb{R}}%
^{(\Delta_{1}+\Gamma_{1},\Delta_{2}+\Gamma_{2})},
\]
where $\left\langle \Delta,\Gamma\right\rangle =\Delta_{1}\Gamma_{1}%
+\Delta_{2}\Gamma_{2}$ and it is easy to prove that this bracket turns
$(\mathrm{End}_{\mathbb{R}}(\Omega_{G}(\mathcal{A}))$ $,[.,.])$ into a
bigraded Lie algebra. This implies, among other things, that
\[
\lbrack\Delta,\Gamma]=-(-1)^{\left\langle \Delta,\Gamma\right\rangle }%
[\Gamma,\Delta]
\]
and the graded Jacobi identity
\begin{equation}
\lbrack\Delta,[\Gamma,\Lambda]]=[[\Delta,\Gamma],\Lambda]+(-1)^{\left\langle
\Delta,\Gamma\right\rangle }[\Gamma,[\Delta,\Lambda]]\text{.} \label{eq2}%
\end{equation}

\begin{definition}
\label{def7}A differential operator on $\Omega_{G}(\mathcal{A})$ of bidegree
$(\Delta_{1},\Delta_{2})$ and order equal to or less than $q$, is an
endomorphism $\Delta\in\mathrm{End}_{\mathbb{R}}^{(\Delta_{1},\Delta_{2}%
)}(\Omega_{G}(\mathcal{A}))$ such that
\[
\lbrack\lbrack...[[\Delta,\mu_{a_{0}}],\mu_{a_{1}}],...],\mu_{a_{q}}]=0
\]
for all $\mu_{a_{i}},i\in\{0,...,q\}$, where $a_{i}\in\Omega_{G}(\mathcal{A})$
and $\mu_{a_{i}}$ denotes the morphism multiplication by $a_{i},\mu_{a_{i}%
}(b)=a_{i}\wedge b$ ($\forall b\in\mathcal{A}$), which has order $0$ and
bidegree that of $a_{i}$. The space of such operators is denoted
$\mathcal{D}if_{q}^{(\Delta_{1},\Delta_{2})}(\Omega_{G}(\mathcal{A}))$ or even
$\mathcal{D}if_{q}^{(\Delta_{1},\Delta_{2})}(\mathcal{A})$ if there is no risk
of confusion.
\end{definition}

We will restrict ourselves to operators vanishing on the identity, that is
$\Delta(\mathbf{1})=0$.

\begin{example}
The first-order differential operators are the derivations: if $\Delta
\in\mathcal{D}if_{1}^{(\Delta_{1},\Delta_{2})}(\mathcal{A}),$ then%
\[
\lbrack\lbrack\Delta,\mu_{a_{0}}],\mu_{a_{1}}]=0
\]
is equivalent to $\Delta(a_{0}\cdot a_{1})=\Delta(a_{0})\cdot a_{1}%
+(-1)^{\left\langle \Delta,a_{0}\right\rangle }a_{0}\cdot\Delta(a_{1}).$ Other
examples are the known insertion operator $i_{A}\in\mathcal{D}if_{p}%
^{(-p,\left\vert A\right\vert )}(\mathcal{A})$ (where $A$ is a graded
$p-$multivector and $\left\vert A\right\vert $ its $\mathbb{Z}-$degree), the
generalized Lie derivative $\mathcal{L}_{A}^{G}=[i_{A},\operatorname{d}%
^{G}]\in\mathcal{D}if_{p}^{(-p+1,\left\vert A\right\vert )}(\mathcal{A})$ and
$\operatorname{d}^{G}\in\mathcal{D}if_{p}^{(1,0)}(\mathcal{A})$.
\end{example}

A useful result states that
\begin{equation}
\lbrack\mathcal{D}if_{q}^{(p_{1},p_{2})}(\mathcal{A}),\mathcal{D}%
if_{q^{\prime}}^{(p_{1}^{\prime},p_{2}^{\prime})}(\mathcal{A})]\subset
\mathcal{D}if_{q+q^{\prime}-1}^{(p_{1}+p_{1}^{\prime},p_{2}+p_{2}^{\prime}%
)}(\mathcal{A}), \label{eq3}%
\end{equation}
and from that formula the following is easily proved:

\begin{proposition}
\label{Prop1}Any differential operator $\Delta\in\mathcal{D}if_{\delta
}^{(\Delta_{1},\Delta_{2})}(\mathcal{A})$ is determined by its action on
graded forms with cohomological degree equal to or less than the order of the
operator, that is, by $\Delta|_{\Omega_{G}^{(p,q)}(\mathcal{A})}$ where $0\leq
p\leq\delta$.
\end{proposition}

\begin{proposition}
\label{Prop2}Any operator $\Delta\in\mathcal{D}if_{q}^{(-q,\left\vert
A\right\vert )}(\mathcal{A})$ has the form $i_{A}$, for some $A\in A_{G}%
^{q}(\mathcal{A})$.
\end{proposition}

This is a consequence of (\ref{eq3}) and the preceding Proposition. Another
very useful algebraic property guarantees that, if $\Delta\in\mathcal{D}%
if_{q}^{(-\Delta_{1},\Delta_{2})}(\mathcal{A})$ with $\Delta_{1}>q$, then
$\Delta\equiv0$. Indeed, by definition%
\[
\Delta:\Omega_{G}^{(p,k)}(\mathcal{A})\rightarrow\Omega_{G}^{(p-\Delta
_{1},k+\Delta_{2})}(\mathcal{A}),
\]
and applying Proposition \ref{Prop1}, $\Delta$ is determined by its action on
$\Omega_{G}^{(p,k)}(\mathcal{A})$ with $0\leq p\leq q$. But for these values,
as $\Delta_{1}>q$, we have $p-\Delta_{1}<0$ and consequently $\Delta=0$.

According to (\ref{eq3}), if we have two graded multivector fields $A\in
A_{G}^{(A_{1},A_{2})}(\mathcal{A}),$ $B\in A_{G}^{(B_{1},B_{2})}(\mathcal{A})
$ then
\[
\lbrack\mathcal{L}_{A}^{G},\iota_{B}]\in\lbrack\mathcal{D}if_{A_{1}}%
^{(-A_{1}+1,A_{2})}(\mathcal{A}),\mathcal{D}if_{B_{1}}^{(-B_{1},B_{2}%
)}(\mathcal{A})]\subset\mathcal{D}if_{A_{1}+B_{1}-1}^{(-(A_{1}+B_{1}%
-1),A_{2}+B_{2})}(\mathcal{A}),
\]
and Proposition \ref{Prop2} enables us to give the following definition:

\begin{definition}
The graded Schouten-Nijenhuis bracket of $A$ and $B$, which will be denoted
$\left[  A,B\right]  _{GSN}$, is the graded multivector field given by
\[
\lbrack\mathcal{L}_{A}^{G},\iota_{B}]=\iota_{\lbrack A,B]_{GSN}}.
\]

\end{definition}

\begin{remark}
Note that $\left[  A,B\right]  _{GSN}\in A_{G}^{(A_{1}+B_{1}-1,A_{2}+B_{2}%
)}(\mathcal{A})$. Also, it is worth to observe that the graded
Schouten-Nijenhuis bracket has some features which are absent in the classical
Schouten-Nijenhuis bracket. For instance, when we take a pair of graded
$2-$vectors $A,B$ one of which has even ghost degree, then $\left[
A,B\right]  _{GSN}=0$ trivially (see \ref{Prop4_1} in Proposition \ref{Prop4} below).
\end{remark}

As an inmediate consequence of the graded Jacobi identity (\ref{eq2}) we have

\begin{proposition}
With the preceding notations
\[
\lbrack\mathcal{L}_{A}^{G},\mathcal{L}_{B}^{G}]=\mathcal{L}_{[A,B]_{GSN}}%
^{G}\text{.}%
\]

\end{proposition}

Also, it is easy to verify other interesting properties of this bracket,
namely: graded anticommutativity, graded Jacobi identity (with which it has
the structure of bigraded Lie algebra) and graded Leibniz rule. The proofs are
simple modifications of those of \cite{Mic 87} and will be omitted.

\begin{proposition}
\label{Prop4}With the preceding notations

\begin{enumerate}
\item \label{Prop4_1}%
\[
\left[  A,B\right]  _{GSN}=-(-1)^{(A_{1}-1)(B_{1}-1)+A_{2}B_{2}}[B,A]_{GSN}.
\]

\item
\begin{align*}
\left[  A,[B,C]_{GSN}\right]  _{GSN}  &  =\left[  [A,B]_{GSN},C\right]
_{GSN}\\
&  +(-1)^{(A_{1}-1)(B_{1}-1)+A_{2}B_{2}}[B,[A,C]_{GSN}]_{GSN}.
\end{align*}

$\ \ \ \ \ \ \ \ \ \ \ \ \ \ \ \ \ \ \ \ $

\item
\[
\left[  A,B\wedge C\right]  _{GSN}=\left[  A,B\wedge C\right]  _{GSN}\wedge
C+(-1)^{(A_{1}-1)B_{1}+A_{2}B_{2}}B\wedge\lbrack A,C]_{GSN}.
\]

\end{enumerate}

Thus, $(A_{G}(\mathcal{A}),\left[  \_,\_\right]  _{GSN})$ has the structure of
a bigraded Gerstenhaber algebra.
\end{proposition}

We can characterize the action of $\left[  \_,\_\right]  _{GSN}$ by giving it
only on the generators of $A_{G}(\mathcal{A})$:%

\begin{align}
&  \left\{
\begin{array}
[c]{l}%
\{0\}=A_{G}^{-1}(\mathcal{A})\ni\left[  a,b\right]  _{GSN}=0,\\
\\
\mathcal{A}=A_{G}^{0}(\mathcal{A})\ni\left[  X,a\right]  _{GSN}=X(a),\\
\\
A_{G}^{1}(\mathcal{A})\ni\left[  X,Y\right]  _{GSN}=\left[  X,Y\right]  ,\\
\end{array}
\right.
\begin{array}
[c]{r}%
\forall a,b\in\mathcal{A}\\
\\
\forall a\in\mathcal{A},\forall X\in A_{G}^{1}(\mathcal{A})\\
\\
\forall X,Y\in A_{G}^{1}(\mathcal{A}),\\
\end{array}
\label{eq4}\\
& \nonumber
\end{align}

\smallskip where $\left[  X,Y\right]  $ denotes the bracket on $\mathrm{End}%
_{\mathbb{R}}(\Omega_{G}(\mathcal{A}))$.

We have mentioned the bigraded Gerstenhaber algebras. Let us recall some
relevant definitions (cfr. \cite{WGS 01}).

\begin{definition}
Let $\mathcal{B}=\sum\limits_{b_{1},b_{2}\in\mathbb{Z}}\mathcal{B}_{b_{1}%
}^{b_{2}}$ be a bigraded commutative algebra over a commutative ring
$\mathcal{R}$. A bigraded Lie algebra structure $[\_,\_]$, of bidegree $(-1,0)
$, on $\mathcal{B}=\sum\limits_{b_{1},b_{2}\in\mathbb{Z}}\mathcal{B}_{b_{1}%
}^{b_{2}}$ is called a bigraded Gerstenhaber algebra bracket if, for each
$B\in\mathcal{B}_{b_{1}}^{b_{2}}$, $[B,\_]$ is a derivation of bidegree
$(b_{1}-1,b_{2})$ of $\mathcal{B}=\sum\limits_{b_{1},b_{2}\in\mathbb{Z}%
}\mathcal{B}_{b_{1}}^{b_{2}}$. In this case, $\mathcal{B}$ is called a
bigraded Gerstenhaber algebra.
\end{definition}

\begin{remark}
A bigraded Gerstenhaber algebra is also called a bigraded Poisson bracket of
bidegree $(-1,0)$. The general definition is the following.
\end{remark}

\begin{definition}
\label{Bipoisson}A bigraded Poisson bracket on $\mathcal{B}$, with bidegree
$\left\vert P\right\vert =(p_{1},p_{2})$, is an $\mathcal{R}-$bilinear map
$\{\_,\_\}:\mathcal{B}\times\mathcal{B}\rightarrow\mathcal{B}$ such that:

\begin{enumerate}
\item $\left\vert \{B_{1},B_{2}\}\right\vert =\left\vert B_{1}\right\vert
+\left\vert B_{2}\right\vert +\left\vert P\right\vert $ , $\forall B_{1}%
,B_{2}\in\mathcal{B}$.

\item $\{B_{1},B_{2}\}=-(-1)^{\left\langle B_{1}+P,B_{2}+P\right\rangle
}\{B_{2},B_{1}\}$.

\item $\{B_{1},B_{2}\cdot B_{3}\}=\{B_{1},B_{2}\}\cdot B_{3}%
+(-1)^{\left\langle B_{1}+P,B_{2}\right\rangle }B_{2}\cdot\{B_{1},B_{3}\}$.

\item $\{B_{1},\{B_{2},B_{3}\}\}=\{\{B_{1},B_{2}\},B_{3}\}+(-1)^{\left\langle
B_{1}+P,B_{2}+P\right\rangle }\{B_{2},\{B_{1},B_{3}\}\}$
\end{enumerate}

$\forall B_{1},B_{2},B_{3}\in\mathcal{B}$.
\end{definition}

\section{The Krasil'shchik-Schouten bracket on supermanifolds}

This Section presents in a condensed form some of the results contained in
\cite{Kra 91}.

Let $(M,\mathcal{A})$ be a supermanifold, with $\mathcal{A}$ a sheaf of graded
commutative algebras over a commutative ring $\mathcal{R}$. We will write
$a\cdot b=(-1)^{ab}b\cdot a$ for the product of sections of $\mathcal{A}$ (the
superfunctions on $(M,\mathcal{A})$).

Now, consider the algebra of multiderivations on $\mathcal{A}$. This is the
bigraded algebra%
\[
\mathcal{D}(\mathcal{A})=\sum\limits_{i\in\mathbb{Z}}\mathcal{D}%
_{i}(\mathcal{A}),
\]
where%
\begin{align*}
\mathcal{D}_{-i}(\mathcal{A})  &  =\{0\}\text{ , }\forall i\in\mathbb{N}\\
\mathcal{D}_{0}(\mathcal{A})  &  =\mathcal{A}\\
\mathcal{D}_{1}(\mathcal{A})  &  =\{F\in\mathrm{Hom}_{\mathcal{R}}%
(\mathcal{A},\mathcal{D}_{0}(\mathcal{A})):F(a\cdot b)=F(a)\cdot
b+(-1)^{aF}a\cdot F(b)\}\\
&  \cdot\cdot\cdot
\end{align*}%
\[%
\begin{tabular}
[c]{r}%
$\mathcal{D}_{i}(\mathcal{A})=\{F\in\mathrm{Hom}_{\mathcal{R}}(\mathcal{A}%
,\mathcal{D}_{i-1}(\mathcal{A})):F(a\cdot b)=F(a)\cdot b+(-1)^{aF}a\cdot
F(b)$\\
$\text{and }(F(a))(b)=-(-1)^{ab}(F(b))(a)\},$%
\end{tabular}
\ \ \ \ \ \
\]

(in $(-1)^{aF}$ we understand the degree of $F$ as a morphism of $\mathbb{Z}-
$graded algebras) and the algebra product is the mapping $\ast:\mathcal{D}%
_{i}(\mathcal{A})\times\mathcal{D}_{j}(\mathcal{A})\rightarrow\mathcal{D}%
_{i+j}(\mathcal{A})$ defined as follows:

\begin{enumerate}
\item If $a,b\in\mathcal{D}_{0}(\mathcal{A})=\mathcal{A}$, then $a\ast
b=a\cdot b$.

\item If $a\in\mathcal{D}_{0}(\mathcal{A}),F\in\mathcal{D}_{1}(\mathcal{A})$,
then $a\ast F\in\mathcal{D}_{1}(\mathcal{A})$ is given by%
\[
a\ast F=a\cdot F=-(-1)^{aF}F\ast a,
\]
where $a\cdot F:\mathcal{A}\rightarrow\mathcal{A}$ acts as $($ $a\cdot F)(b)=
$ $a\cdot(F(b))$, $\forall b\in\mathcal{A}$.

\item Recursively, if $F\in\mathcal{D}_{f}(\mathcal{A}),G\in\mathcal{D}%
_{g}(\mathcal{A})$, then $F\ast G\in\mathcal{D}_{f+g}(\mathcal{A})$ acts as%
\begin{equation}
(F\ast G)(a)=F\ast G(a)+(-1)^{aG+g}F(a)\ast G\text{ , }\forall a\in
\mathcal{A}. \label{eq5}%
\end{equation}

\end{enumerate}

\begin{remark}
Clearly, for each $i\in\mathbb{Z}$, $\mathcal{D}_{i}(\mathcal{A})$ is an
$\mathcal{A}-$module. By definition, given $a\in\mathcal{A}$ we have
$F(a)\in\mathcal{D}_{f-1}(\mathcal{A}),G(a)\in\mathcal{D}_{g-1}(\mathcal{A})$,
so (assuming the product $\ast$ defined on $\mathcal{D}_{f-1}(\mathcal{A}%
)\times\mathcal{D}_{g}(\mathcal{A})$ and $\mathcal{D}_{f}(\mathcal{A}%
)\times\mathcal{D}_{g-1}(\mathcal{A})$), formula (\ref{eq5}) yields an element
of $\mathcal{D}_{f+g}(\mathcal{A})$.
\end{remark}

\begin{proposition}
The product $\ast:\mathcal{D}_{i}(\mathcal{A})\times\mathcal{D}_{j}%
(\mathcal{A})\rightarrow\mathcal{D}_{i+j}(\mathcal{A})$ is associative,
distributive with respect to addition and it verifies%
\[
F\ast G=(-1)^{FG+fg}G\ast F
\]
when $F\in\mathcal{D}_{f}(\mathcal{A}),G\in\mathcal{D}_{g}(\mathcal{A})$.
\end{proposition}

Note that $\mathcal{D}(\mathcal{A})=\sum\limits_{i\in\mathbb{Z}}%
\mathcal{D}_{i}(\mathcal{A})$ has a grading inherited from that of
$\mathcal{A}$. Moreover, we can consider on it another additional degree,
given by the $f\in\mathbb{Z}$ to which an $F\in\mathcal{D}(\mathcal{A}%
)=\sum\limits_{f\in\mathbb{Z}}\mathcal{D}_{f}(\mathcal{A})$ belongs, thus
getting what is called a partitioned bigraded algebra with degree indexes
$(f,F)$ and marked degree $f$. So, we will write $\left\vert F\right\vert
=(f,F)$ when $F\in\mathcal{D}_{f}(\mathcal{A})$.

Let us explain why the elements of $\mathcal{D}(\mathcal{A})$ are called
multiderivations. This result will be important later.

\begin{proposition}
For each $i\in\mathbb{Z}$, there exists an isomorphism%
\[
\varphi:A_{G}^{i}(\mathcal{A})=\mathrm{Hom}_{\mathcal{A}}(\Omega_{G}%
^{i}(\mathcal{A}),\mathcal{A})\rightarrow\mathcal{D}_{i}(\mathcal{A}).
\]

\end{proposition}

\begin{proof}
First, note that%
\begin{align*}
\mathcal{D}_{1}(\mathcal{A})  &  =\{F\in\mathrm{Hom}_{\mathcal{R}}%
(\mathcal{A},\mathcal{A}):F(a\cdot b)=F(a)\cdot b+(-1)^{ab}a\cdot F(b)\}\\
&  =\mathrm{Der}(\mathcal{A})=A_{G}^{1}(\mathcal{A}).
\end{align*}
Let us assume $\mathcal{D}_{i}(\mathcal{A})\simeq A_{G}^{i}(\mathcal{A})$ for
$i\in\mathbb{N}$; then, if $F\in\mathcal{D}_{i+1}(\mathcal{A})$ we have that
$F:\mathcal{A}\rightarrow\mathcal{D}_{i}(\mathcal{A})$ verifies%
\[
F(a\cdot b)=F(a)\cdot b+(-1)^{aF}a\cdot F(b),
\]
so $F\in\mathcal{D}_{1}(\mathcal{A})\otimes\mathcal{D}_{i}(\mathcal{A})\simeq
A_{G}^{1}(\mathcal{A})\otimes A_{G}^{i}(\mathcal{A})$. Moreover, \ (by
definition of $\mathcal{D}_{i}(\mathcal{A})$),%
\[
(F(a))(b)=-(-1)^{ab}(F(b))(a),
\]
thus, actually $F\in A_{G}^{1}(\mathcal{A})\wedge A_{G}^{i}(\mathcal{A})$.

Reciprocally, if $A\wedge B\in A_{G}^{i+1}(\mathcal{A})=A_{G}^{1}%
(\mathcal{A})\wedge A_{G}^{i}(\mathcal{A})$ is a decomposable graded
multivector field, we define the mapping%
\[
F_{A\wedge B}:\mathcal{A}\rightarrow\mathcal{D}_{i}(\mathcal{A})\simeq
A_{G}^{i}(\mathcal{A})
\]
through%
\[
F_{A\wedge B}(a)=A\wedge B(a)+(-1)^{ab}A(a)\cdot B,
\]
which verifies (this is an easy computation) $F_{A\wedge B}(a\cdot
b)=F(a)\cdot b+(-1)^{a(A+B)}a\cdot F_{A\wedge B}(b)$ and $(F_{A\wedge
B}(a))(b)=-(-1)^{ab}(F_{A\wedge B}(b))(a)$.

Finally, note that if we start from $A\wedge B\in A_{G}^{i}(\mathcal{A})$,
consider $F_{A\wedge B}\in\mathcal{D}_{i}(\mathcal{A})$ and then return to
$A_{G}^{i}(\mathcal{A})$ applying the preceding procedure, we will arrive at
the element $A\wedge B$. The same thing happens following the reverse path, so
we have an isomorphism $A_{G}^{i}(\mathcal{A})\simeq\mathcal{D}_{i}%
(\mathcal{A}).$
\end{proof}

\begin{remark}
As a consequence of this result, note that the marked degree in $\mathcal{D}%
(\mathcal{A})=A_{G}(\mathcal{A})$ is the cohomological degree.
\end{remark}

In addition to the product $\ast$, we can define another structure on
$\mathcal{D}(\mathcal{A})$, a bracket $[\![\_,\_]\!]$ called the
Krasil'shchik-Schouten bracket. It will be constructed recursively in such a
way that $[\![\_,\_]\!]:\mathcal{D}_{i}(\mathcal{A})\times\mathcal{D}%
_{j}(\mathcal{A})\rightarrow\mathcal{D}_{i+j-1}(\mathcal{A})$.

\begin{enumerate}
\item If $a,b\in\mathcal{D}_{0}(\mathcal{A})=\mathcal{A}$, then%
\begin{equation}
\lbrack\![a,b]\!]=0. \label{eq6}%
\end{equation}

\item If $a\in\mathcal{D}_{0}(\mathcal{A}),F\in\mathcal{D}_{1}(\mathcal{A})$,
then%
\begin{equation}
\lbrack\![F,a]\!]=F(a)=(-1)^{aF+f}[\![a,F]\!]. \label{eq7}%
\end{equation}

\item Recursively, if $F\in\mathcal{D}_{f}(\mathcal{A}),G\in\mathcal{D}%
_{g}(\mathcal{A})$, then the action of $[\![F,G]\!]:\mathcal{A}\rightarrow
\mathcal{D}_{f+g-2}(\mathcal{A})$ is given by%
\begin{equation}
\lbrack\![F,G]\!](a)=[\![F,G(a)]\!]+(-1)^{FG+g}[\![F(a),G]\!]\text{ , }\forall
a\in\mathcal{A}\text{.} \label{eq8}%
\end{equation}

\end{enumerate}

The main properties of $[\![\_,\_]\!]$ are gathered in the following result
(due to Krasil'shchik, \cite{Kra 91}).

\begin{proposition}
\label{Leibniz}The mapping $[\![\_,\_]\!]:\mathcal{D}_{i}(\mathcal{A}%
)\times\mathcal{D}_{j}(\mathcal{A})\rightarrow\mathcal{D}_{i+j-1}%
(\mathcal{A})$ is $\mathcal{R}-$bilinear and it verifies:

\begin{enumerate}
\item
\[
\lbrack\![F,G]\!]=-(-1)^{FG+(f-1)(g-1)}[\![G,F]\!].
\]

\item \label{item2}%
\[
\lbrack\![F,G\ast H]\!]=[\![F,G]\!]\ast H+(-1)^{FG+(f-1)g}G\ast\lbrack
\![F,H]\!].
\]

\item
\[
\lbrack\![F,[\![G,H]\!]]\!]=[\![[\![F,G]\!],H]\!]+(-1)^{FG+(f-1)(g-1)}%
[\![G,[\![F,H]\!]]\!].
\]
$\forall F\in\mathcal{D}_{f}(\mathcal{A}),G\in\mathcal{D}_{g}(\mathcal{A}%
),H\in\mathcal{D}_{h}(\mathcal{A})$.
\end{enumerate}
\end{proposition}

Taking into account that $\mathcal{D}(\mathcal{A})$ is a marked bigraded
algebra, we see that $[\![\_,\_]\!]$ is an example of marked bigraded Poisson
bracket on $\mathcal{D}(\mathcal{A})$, with marked bidegree $(-1,0)$ (the $-1$
takes place in the marked index), that is to say, $\left(  \mathcal{D}%
(\mathcal{A}),[\![\_,\_]\!]\right)  $ is a bigraded Gerstenhaber algebra.

The Krasil'shchik-Schouten bracket is a tool which unifies many of the objects
encountered in the theory of calculus over modules (this was indeed one of the
reasons Krasil'shchik introduced it in \cite{Kra 91}). For multivectors on
supermanifolds it coincides with the graded Schouten-Nijenhuis bracket of
Paragraph \ref{GSN} (see the next Example), but the construction of
Krashil'shchick is far more general and it applies to the case where
$\mathcal{A}$ is a sheaf of $k-$graded commutative algebras.

\begin{example}
[the graded Schouten-Nijenhuis bracket]\label{Example}Consider $M\supset
U\mapsto\mathcal{A}(U)$ a sheaf of graded commutative algebras. We know that
the sheaves of multiderivations on $\mathcal{A}$, $\mathcal{D}(\mathcal{A})$,
are naturally isomorphic to the sheaves of graded multivectors $A_{G}%
(\mathcal{A})$, where we have defined the graded Schouten-Nijenhuis bracket
$[\_,\_]_{GSN}$ (see \ref{GSN}). This isomorphism translates $[\![\_,\_]\!]$
into $[\_,\_]_{GSN}$, more precisely
\end{example}

\begin{proposition}
$\varphi:(A_{G}^{i}(\mathcal{A}),[\_,\_]_{GSN})\rightarrow(\mathcal{D}%
_{i}(\mathcal{A}),[\![\_,\_]\!])$ is an isomorphism of bigraded Gerstenhaber algebras.
\end{proposition}

\begin{proof}
To see this, note that the expressions (\ref{eq6}), (\ref{eq7}), (\ref{eq8})
reduce to (\ref{eq4}) when we consider the generators of $A_{G}^{i}%
(\mathcal{A})$: $a\in\mathcal{A}=A_{G}^{0}(\mathcal{A})$ and $X\in A_{G}%
^{1}(\mathcal{A})$.
\end{proof}

\begin{example}
In particular, if we choose $\mathcal{A}(U)=C^{\infty}(U)$ (that is, the
trivial sheaf of differentiable functions on $M$), then $\mathcal{D}%
(C^{\infty}(U))\simeq A(C^{\infty}(U))=\Lambda_{C^{\infty}(U)}(TU)$, that is,
multiderivations are usual multivectors, and $[\![\_,\_]\!]$ is nothing but
the usual Schouten-Nijenhuis bracket.
\end{example}

\section{Divergences and generators of the Krashil'shchick-Schouten bracket}

As we mentioned before, the Krasil'shchik-Schouten bracket can be viewed as a
(marked bigraded) Poisson bracket on $\mathcal{D}(\mathcal{A})$, so it makes
sense to ask for its generating operators and how to construct them. Let us
recall the basic concepts.

\begin{definition}
Let $(\mathcal{B},\{\_,\_\})$ be a bigraded Poisson structure, that is,
$\{\_,\_\}$ a bigraded Poisson bracket as in Definition \ref{Bipoisson} with
bidegree $P$. A second-order differential operator on $\mathcal{B}$,
$\Delta\in\mathcal{D}if_{2}(\mathcal{B}),$ is called a generating operator for
the bracket $\{\_,\_\}$ if, for all $B_{1},B_{2}\in\mathcal{B}$, we can
express $\{B_{1},B_{2}\}$ as%
\begin{equation}
\{B_{1},B_{2}\}=(-1)^{\left\langle B_{1},P\right\rangle }(\Delta(B_{1}\cdot
B_{2})-\Delta(B_{1})\cdot B_{2}-(-1)^{\left\langle B_{1},P\right\rangle }%
B_{1}\cdot\Delta(B_{2}))\text{.} \label{eq8b}%
\end{equation}

\end{definition}

In the following, we will consider the case $\mathcal{B}=\mathcal{D}%
(\mathcal{A})=A_{G}(\mathcal{A})$ with the algebra structure given by the
product $\wedge$ on graded multivectors. We show that, as in the classical
theory, generating operators of the bigraded Gerstenhaber structure on
$\mathcal{D}(\mathcal{A})$ can be obtained with the aid of divergence operators.

\begin{definition}
A divergence operator on $\mathcal{A}$ is a linear morphism of sheaves of
$\mathbb{R}-$vector spaces%
\begin{equation}%
\begin{tabular}
[c]{r}%
$\mathrm{div}:\mathrm{Der}(\mathcal{A})\rightarrow\mathcal{A}$\\
$D\mapsto\mathrm{div}(D)$%
\end{tabular}
\ \ \ \ \label{eq9}%
\end{equation}
such that $\forall a\in\mathcal{A}$,%
\[
\mathrm{div}(a\cdot D)=a\cdot\mathrm{div}(D)+(-1)^{aD}D(a)\text{.}%
\]

\end{definition}

Examples of divergence operators can be obtained from the Berezinian sheaf of
a supermanifold and their associated Berezinian volume elements, or from
graded connections. We refer the reader to \cite{Kos-Mon 02} for details.

The main result is then the following.

\begin{theorem}
Let $\mathrm{div}$ be a divergence operator on $\mathcal{A}$. Then,
$-\mathrm{div}$ can be uniquely extended into a generating operator
$\Delta_{\mathrm{div}}$ of bidegree $(-1,0)$ of the bigraded Gerstenhaber
algebra $(A_{G}(\mathcal{A})=\Lambda_{\mathcal{A}}\mathrm{Der}(\mathcal{A}%
),[\_,\_]_{GSN})$, and $\Delta_{\mathrm{div}}$ commutes (in the graded sense)
with the insertion of the graded $1-$form $d^{G}a$, $\iota_{d^{G}a}$, for any
$a\in\mathcal{A}$.
\end{theorem}

\begin{proof}
Let us write $\Delta_{\mathrm{div}}$ simply as $\Delta$. First, note that for
any $a\in\mathcal{A}$ and graded multivectors $A,B\in A_{G}(\mathcal{A}) $:%
\begin{equation}
\iota_{d^{G}a}\left(  A\wedge B\right)  =\iota_{d^{G}a}\left(  A\right)
\wedge B+(-1)^{A_{1}+A_{2}a}A\wedge\iota_{d^{G}a}\left(  B\right)
\label{eq10}%
\end{equation}
and%
\begin{equation}
\iota_{d^{G}a}\left[  A,B\right]  _{GSN}=\left[  \iota_{d^{G}a}A,B\right]
_{GSN}+(-1)^{A_{1}-1+A_{2}a}\left[  A,\iota_{d^{G}a}B\right]  _{GSN}.
\label{eq11}%
\end{equation}
Now, because its cohomological degree must be $-1$, the operator $\Delta$ has
to vanish on $A_{G}^{0}(\mathcal{A})\simeq\mathcal{A}$ and, because we want it
to be an extension, it has to coincide with $-\mathrm{div}$ on $A_{G}%
^{1}(\mathcal{A})\simeq\mathrm{Der}(\mathcal{A})$:%
\[%
\begin{array}
[c]{l}%
\Delta|_{A_{G}^{0}(\mathcal{A})}\doteq0\\
\Delta|_{A_{G}^{1}(\mathcal{A})}\doteq-\mathrm{div.}%
\end{array}
\]
Assume that it is defined on graded multivectors $A$ of bidegree $(A_{1}%
,A_{2})$ with cohomological degree $A_{1}\leq k$, that%
\begin{equation}
\iota_{d^{G}a}(\Delta(A))=-\Delta(\iota_{d^{G}a}A) \label{eq12}%
\end{equation}
for all $a\in\mathcal{A}$ and $A\in A_{G}(\mathcal{A})$ with $A_{1}\leq k$,
and that $\Delta(A)$ satisfies the defining relation (\ref{eq8b}) for
generators of the bracket $[\_,\_]_{GSN}$ for all graded multivectors $A$ and
$B$ such that $A_{1}+B_{1}\leq k$ (sum of cohomological degrees). Relation
(\ref{eq12}) then defines $\Delta(C)$ for graded multivectors $C$ of
cohomological degree $k+1$. Let us prove that (\ref{eq8b}) is satisfied for
$A_{1}+B_{1}\leq k+1$.

Indeed, let $A,B\in A_{G}(\mathcal{A})$ be as above ($A_{1}+B_{1}\leq k$) and
let $X\in A_{G}^{1}(\mathcal{A})$, so $A\wedge X\wedge B$ has cohomological
degree $k+1$. Using (\ref{eq11}) we write%
\[
\iota_{d^{G}a}\left[  A,X\wedge B\right]  _{GSN}=\left[  \iota_{d^{G}%
a}A,X\wedge B\right]  _{GSN}+(-1)^{A_{1}-1+A_{2}a}\left[  A,\iota_{d^{G}%
a}\left(  X\wedge B\right)  \right]  _{GSN}.
\]
Let us study the right hand side. Since the sum of the cohomological degrees
of $\iota_{d^{G}a}A$ and $X\wedge B$ is $k$, applying the induction
hypothesis:%
\begin{align*}
\left[  \iota_{d^{G}a}A,X\wedge B\right]  _{GSN}  &  =(-1)^{A_{1}-1}\left(
\Delta\left(  \iota_{d^{G}a}A\wedge X\wedge B\right)  \right. \\
&  \left.  -\Delta\left(  \iota_{d^{G}a}A\right)  \wedge X\wedge
B-(-1)^{A_{1}-1}\iota_{d^{G}a}A\wedge\Delta\left(  X\wedge B\right)  \right)
,
\end{align*}
and%
\begin{align*}
(-1)^{A_{1}-1+A_{2}a}\left[  A,\iota_{d^{G}a}\left(  X\wedge B\right)
\right]  _{GSN}  &  =-(-1)^{A_{2}a}\left\{  \Delta\left(  A\wedge\iota
_{d^{G}a}\left(  X\wedge B\right)  \right)  \right. \\
&  -\Delta\left(  A\right)  \wedge\iota_{d^{G}a}\left(  X\wedge B\right) \\
&  \left.  -(-1)^{A_{1}}A\wedge\Delta\left(  \iota_{d^{G}a}\left(  X\wedge
B\right)  \right)  \right\}  ,
\end{align*}
so we get, using (\ref{eq12}):%
\begin{align*}
\iota_{d^{G}a}\left[  A,X\wedge B\right]  _{GSN}  &  =(-1)^{A_{1}-1}%
\Delta\left(  \iota_{d^{G}a}A\wedge X\wedge B+(-1)^{A_{1}+A_{2}a}A\wedge
\iota_{d^{G}a}\left(  X\wedge B\right)  \right) \\
&  -(-1)^{A_{1}}\iota_{d^{G}a}\Delta\left(  A\right)  \wedge X\wedge
B-\iota_{d^{G}a}A\wedge\Delta\left(  X\wedge B\right) \\
&  +(-1)^{A_{2}a}\Delta\left(  A\right)  \wedge\iota_{d^{G}a}\left(  X\wedge
B\right) \\
&  -(-1)^{A_{1}+A_{2}a}A\wedge\iota_{d^{G}a}\Delta\left(  X\wedge B\right)  ,
\end{align*}
thus, by (\ref{eq10}) and (\ref{eq12}) again:%
\begin{align*}
\iota_{d^{G}a}\left[  A,X\wedge B\right]  _{GSN}  &  =(-1)^{A_{1}-1}%
\Delta\left(  \iota_{d^{G}a}\left(  A\wedge X\wedge B\right)  \right) \\
&  +(-1)^{A_{1}-1}\iota_{d^{G}a}\left(  \Delta\left(  A\right)  \wedge X\wedge
B\right) \\
&  -\iota_{d^{G}a}\left(  A\wedge\Delta\left(  X\wedge B\right)  \right) \\
&  =(-1)^{A_{1}}\iota_{d^{G}a}\left(  \Delta\left(  A\wedge X\wedge B\right)
\right. \\
&  -\Delta\left(  A\right)  \wedge X\wedge B\\
&  \left.  -(-1)^{A_{1}}A\wedge\Delta\left(  X\wedge B\right)  \right)  .
\end{align*}
That is, for all $a\in\mathcal{A}:$%
\begin{align*}
\iota_{d^{G}a}\left[  A,X\wedge B\right]  _{GSN}  &  =\iota_{d^{G}a}\left(
(-1)^{A_{1}}\Delta\left(  A\wedge X\wedge B\right)  \right. \\
&  -\Delta\left(  A\right)  \wedge X\wedge B\\
&  \left.  -(-1)^{A_{1}}A\wedge\Delta\left(  X\wedge B\right)  \right)  ,
\end{align*}
or%
\[
\left[  A,X\wedge B\right]  _{GSN}=(-1)^{A_{1}}\left(  \Delta\left(  A\wedge
X\wedge B\right)  -\Delta\left(  A\right)  \wedge X\wedge B-(-1)^{A_{1}%
}A\wedge\Delta\left(  X\wedge B\right)  \right)  .
\]

\end{proof}

\begin{corollary}
Let $\mathrm{div}$ be a divergence operator on $\mathcal{A}$. Then,
$-\mathrm{div}$ can be uniquely extended into a generating operator
$\Delta_{\mathrm{div}}$ of bidegree $(-1,0)$, of the bigraded Gerstenhaber
algebra $\mathcal{D}(\mathcal{A})$ endowed with the Krashil'shchick-Schouten
bracket $[\![\_,\_]\!]$.
\end{corollary}

\begin{proof}
This follows from the isomorphism defined in Example \ref{Example}.
\end{proof}

\section{Some comments on the Jacobi case}

In a very recent paper, J. Grabowski and G. Marmo (see \cite{Gra-Mar 03}) have
extended the Krasil'shchik-Schouten bracket to a $(n+1)-$graded Jacobi
bracket, this time defined on the commutative algebras of multidifferential
operators of first order (notice the change in the terminology: they speak
about polydifferential operators). The graded multiderivations we are
considering in this paper are a graded subspace of that of multidifferential
operators\footnote{See equation $(92)$ in \cite{Gra-Mar 03}. Note that they
use $\mathcal{D}^{\alpha} $ not for multiderivations, as we do, but for
multidifferential operators.} of first order, and are characterized by the
fact that they kill the unit element, that is, $D(\mathbf{1})=0$ for any
graded multiderivation. The direct consequence of this fact is that, for the
$(n+1)-$graded Jacobi bracket, the Leibniz rule (item (\ref{item2}) in
Proposition \ref{Leibniz}) is replaced by a generalized expression such as%
\begin{equation}
\lbrack\![F,G\ast H]\!]=[\![F,G]\!]\ast H+(-1)^{FG+(f-1)g}G\ast\lbrack
\![F,H]\!]-[\![F,\mathbf{1}]\!]\ast H\ast G. \label{eq13}%
\end{equation}

Now, the question whether the $m-$graded Jacobi brackets admit generating
operators in the same way that the Krasil'shchik-Schouten bracket does, arises
naturally. This is a very interesting question but it can not be answered by a
straightforward generalization of the techniques used in this paper. This is
easy to see even in the case considered by Koszul in \cite{Koz 85} (the
Schouten-Nijenhuis bracket), a particular instance of Krasil'shchik's construction.

Indeed, let $\mathcal{A}=\bigoplus\limits_{p\geq0}\mathcal{A}^{p}$ be a
$\mathbb{Z}_{2}-$graded commutative algebra, so $ab=(-1)^{ab}ba$ for each
$a\in\mathcal{A}^{a},b\in\mathcal{A}^{b}$, and consider the $\mathbb{Z}_{2}-
$graded Lie algebra of endomorphisms $\mathrm{End}(\mathcal{A})=\bigoplus
\limits_{k\in\mathbb{Z}}\mathrm{End}^{k}(\mathcal{A})$, with the bracket as in
Paragraph \ref{GSN}. Koszul defines, for each endomorphism $D$ and
$r\in\mathbb{Z}$, an $r-$form on $\mathcal{A}$ with values on $\mathcal{A}$,
$\Phi_{D}^{r}(a_{1},...,a_{r})$, such that for instance:%
\begin{align*}
\Phi_{D}^{1}(a)  &  =D(a)-D(\mathbf{1})a,\\
\Phi_{D}^{2}(a,b)  &  =D(ab)-D(a)b-(-1)^{aD}aD(b)+D(\mathbf{1})ab,\\
\Phi_{D}^{3}(a,b,c)  &  =D(abc)-D(a)bc-(-1)^{aD}aD(bc)-(-1)^{b(a+D)}bD(ac)\\
&  +D(a)bc+(-1)^{aD}aD(b)c+(-1)^{D(a+b)}abD(c)-D(\mathbf{1})abc,
\end{align*}
and it is easily verified that the following relation holds:%
\begin{equation}
\Phi_{D}^{3}(a,b,c)=\Phi_{D}^{2}(a,bc)-\Phi_{D}^{2}(a,b)c-(-1)^{bc}\Phi
_{D}^{2}(a,c)b. \label{eq14}%
\end{equation}
This is the embryo of the Leibniz rule for the Schouten-Nijenhuis bracket.
Note that it is independent of the property $D(\mathbf{1})=0$.

The connection with the results presented here, rests upon realizing that the
definition of $\Phi_{D}^{k}$ can be given in terms of the commutator of
endomorphisms (with the notations of Definition \ref{def7}) as%
\[
\Phi_{D}^{k}(a_{1},...,a_{k})=[[[...[D,\mu_{a_{1}}],...],\mu_{a_{k-1}}%
],\mu_{a_{k}}](\mathbf{1}).
\]

Now, given a second-order differential operator $D$ with odd degree
$\left\vert D\right\vert $, Koszul defines the associated bracket on
$\mathcal{A}$, $[\_,\_]_{D}$,%
\[
\lbrack a,b]_{D}=(-1)^{a}\Phi_{D}^{2}(a,b),\text{ }\forall a,b\in\mathcal{A},
\]
and it is immediate, from the explicit expressions for $\Phi_{D}^{2}(a,b)$
above, that it satisfies%
\begin{equation}
\lbrack a,b]_{D}=-(-1)^{(a-D)(b+D)}[b,a]_{D} \label{eq15}%
\end{equation}
and%
\begin{equation}
\lbrack a,bc]_{D}=[a,b]_{D}c+(-1)^{(a+D)b}b[a,c]_{D}, \label{eq16}%
\end{equation}
the last property being a consequence of (\ref{eq14}) and the fact that $D$ is
second-order (so $\Phi_{D}^{3}(a,b,c)=0$). If, in addition, $D(\mathbf{1})=0$
and $D^{2}$ is second-order, then $[\_,\_]_{D}$ also satisfies the Jacobi identity.

A bracket on $\mathcal{A}$ is said to have generating operator $D$ if it is of
the type $[\_,\_]_{D}$. But, as we see from (\ref{eq16}), such a bracket
derived from $\Phi_{D}^{2}$ always gives a \textquotedblleft Leibniz
rule\textquotedblright, and not a generalized expression such as (\ref{eq13}).
Thus, in order to speak about generating operators for $n-$graded Jacobi
brackets along these lines, another (different) approach is needed.

However, there is still another way in which one can think of generating
operators for $n-$graded Jacobi structures, and it has its foundations in some
results presented in \cite{Gra-Mar 03}: concretely Theorem $12$ and Corollary
$5$ in that reference, where it is proved that the space of multidifferential
operators of first order can be decomposed into direct sums of subspaces of multiderivations.

For the case we are considering in this paper, where $\mathcal{A}$ is a graded
$1-$algebra, the results of Grabowski and Marmo state the decomposition (in
our notation!)%
\[
\mathcal{D}if_{q}(\mathcal{A})=\mathcal{D}_{q}(\mathcal{A})\oplus\cdot
\cdot\cdot\oplus\mathcal{D}_{0}(\mathcal{A}),
\]
where the Krasil'shchik-Schouten bracket $[\![\_,\_]\!]$ is defined on each
$\mathcal{D}_{i}(\mathcal{A})$, $0\leq i\leq q$. Then, applying Lemma $4$ in
\cite{Gra-Mar 03} we can extend this bracket to $\mathcal{D}if_{q}%
(\mathcal{A})$. In this context, it makes sense to say that a generating
operator for the Krasil'shchik-Schouten bracket is also a generating operator
for the extended Jacobi-Krasil'shchik-Schouten bracket introduced by Grabowski
and Marmo.

\end{document}